\documentclass[review]{elsarticle}

\PassOptionsToPackage{hyphens}{url}
\usepackage{hyperref}
\usepackage{multirow}
\pdfoutput=1

\makeatletter
\def\ps@pprintTitle{%
  \let\@oddhead\@empty
  \let\@evenhead\@empty
  \let\@oddfoot\@empty
  \let\@evenfoot\@oddfoot
}
\makeatother

\usepackage{color}
\usepackage{amsmath}
\usepackage{makecell}
\usepackage{pdflscape}
\usepackage{breqn}
\usepackage{booktabs}
\usepackage{longtable}

\begin{document}

\begin{frontmatter}


\title{Improving Distribution System Resilience\\ by Undergrounding Lines and\\ Deploying Mobile Generators}

\author{Babak~Taheri}
\author{Daniel~K.~Molzahn}
\author{Santiago~Grijalva} 

\address{\textit{School of Electrical and Computer Engineering} \\
\textit{Georgia Institute of Technology}\\
{taheri@gatech.edu, molzahn@gatech.edu, sgrijalva@ece.gatech.edu}}%
\address{Atlanta, GA, USA}



\cortext[]{Corresponding author: Daniel K. Molzahn, Email: \textcolor{blue}{molzahn@gatech.edu}}


\begin{abstract}
To improve the resilience of electric distribution systems, this paper proposes a stochastic multi-period mixed-integer linear programming model that determines where to underground distribution lines and how to coordinate mobile generators in order to serve critical loads during extreme events. The proposed model represents the service restoration process using the linearized DistFlow approximation of the AC power flow equations as well as binary variables for the undergrounding statuses of the lines, the configurations of switches, and the locations of mobile generators during each time period. The model also enforces a radial configuration of the distribution network and considers the transportation times needed to reposition the mobile generators. Using an extended version of the IEEE 123-bus test system, numerical simulations show that combining the ability to underground distribution lines with the deployment of mobile generators can significantly improve the resilience of the power supply to critical loads.
\end{abstract}


\end{frontmatter}


\section{Introduction}
\label{sec:Introduction}


Natural disasters have caused large-scale power
outages in recent years, with the total cost of 308 major natural disasters since 1980 exceeding \$2 trillion in the United States alone~\cite{NOAA}. 
Accordingly, the resilience of distribution systems has become a topic of substantial interest for researchers and distribution utilities.
The National Infrastructure Advisory Council (NIAC) states that resilience is ``the ability to reduce the magnitude and/or duration of disruptive events. The effectiveness of a resilient infrastructure or enterprise depends upon its ability to anticipate, absorb, adapt to, and/or rapidly recover from a potentially disruptive event'' \cite{berkeley2010framework}. In 2011, the UK Energy Research Center provided a similar definition for the resilience as ``the capacity of an energy system to tolerate disturbance and to continue to deliver affordable energy services to consumers. A resilient energy system can speedily recover from shocks and can provide alternative means of satisfying energy service needs in the event of changed external circumstances'' \cite{chaudry2011building}. Survivability and swift restoration capabilities are the key features of resilient distribution systems.

Resilience can be improved by reducing the initial impact of a disaster and quickly restoring the supply of power after a disaster occurs. 
In the literature, two different approaches have been proposed to enhance distribution system resilience, namely, infrastructure hardening and smart operational strategies~\cite{gholami2018toward}. Hardening strategies reduce the impacts of disasters. For example, the approach in \cite{taheri2019hardening} identifies optimal locations to harden lines and place switches in order to reduce the effects of high-impact low-probability (HILP) events and enhance the restoration performance of the system.  As another example, the authors of \cite{ma2019resilience} enhance distribution system resilience by combining line hardening, installing distributed generators, and adding remote-controlled switches. 

On the other hand, smart operational strategies mainly
focus on the rapid recovery of distribution systems by reconfiguring the
network and using available resources to provide energy to disconnected loads 
after an extreme event. 
As an example of smart operational strategies, our previous work in \cite{taheri2019distribution} proposes a three-stage service restoration model for improving distribution system resilience using pre-event reconfiguration and post-event restoration considering both remote-controlled switches and manual switches. Additionally, the authors of \cite{Amirioun2018Resilience} propose a proactive management scheme for microgrids to cope with severe windstorms. This scheme minimizes the number of vulnerable branches in service while the total loads are supplied. Also, the study in \cite{Lei2019Resilient} presents a two-stage stochastic programming method for the optimal scheduling of microgrids in the face of HILP events considering uncertainties associated with wind generation, electric vehicles, and real-time market prices. For further examples of both infrastructure hardening and smart operational strategies, see~\cite{bhusal2020,jufri2019,mahzarnia2020}. 

Recent research has also studied distribution system restoration problems that consider the deployment of mobile generators~\cite{Taheri2021,Amirioun2018Resilience,Gholami2016Microgrid,Lei2019Resilient,dugan2021application}. In \cite{Taheri2021}, 
we proposed a two-stage strategy involving 1)~a preparation stage that reconfigures the distribution system and pre-positions repair crews and mobile generators and 2)~a post-HILP stage that 
solves a stochastic mixed-integer linear programming (MILP) model to restore the system using distributed generators, mobile generators, and reconfiguration of switches. 
Building on this existing work, this paper proposes a resilience enhancement strategy that aims to supply power to critical loads by selectively adding new underground distribution lines prior to a disaster while considering the capabilities of mobile generators to restore power during the initial hours of a disaster. We will show that long-term line undergrounding decisions which are cognizant of short-term mobile generator deployments yield superior results relative to undergrounding decisions made without considering mobile generators. We next motivate the use of underground lines and mobile generators for improving distribution system resilience.

\textbf{\textit{Connecting Critical Loads by Underground Lines}}: Critical loads, such as hospitals, water pumping facilities, and emergency shelters, are prioritized for energization after extreme events. Improving the connectivity of critical loads can increase the resilience of their power supply. There are various advantages and disadvantages to adding new connections via overhead versus underground distribution lines. For example, compared to overhead distribution lines, underground lines provide reduced likelihood of damage during natural disasters and improved aesthetics~\cite{liu2020resilient}. However, undergrounding all of the distribution lines can be prohibitively expensive and also has drawbacks such as utility employee work hazards during faults and manhole inspections as well as potential susceptibility to flooding and storm surges. To take advantage of the reduced susceptibility of underground distribution lines to extreme events while avoiding their extra costs and potential downsides, we focus on adding new underground lines that support power supply to critical loads. Our proposed strategy determines which lines to underground in order to improve resilience over a range of disaster scenarios while considering the capabilities of mobile generators during the restoration process. 

\textbf{\textit{Mobile Generators}}: Mobile generators are increasingly being used by utility companies~\cite{Dominion_energy, SCE, byd}. Careful deployments of mobile generators can significantly enhance the resilience of distribution systems~\cite{dugan2021application}. In this paper, mobile generators are employed to improve the restoration process in the aftermath of HILP events. Since they can be connected to various locations, mobile generators provide significant flexibility for responding to an extreme event.



To summarize, the need to improve distribution system resilience motivates stronger connections between critical loads as well as more reliable and flexible power sources. When utilized appropriately, underground distribution lines and mobile generators can address these needs. To this end, the main contribution of this paper is a multiperiod model for distribution system restoration that considers the ability to underground lines, reconfigure the network, and reposition mobile generators in order to serve critical loads. 

The remainder of this paper is organized as follows: Section~\ref{sec:Mathematical Formulation}
describes mathematical formulations of the stochastic MILP model. To demonstrate this model, Section~\ref{sec:Results and Discussion} presents a case study and simulation results. Finally, Section~\ref{sec:Conclusion} concludes the paper and discusses future work.


\section{Mathematical Formulation}
\label{sec:Mathematical Formulation}
This section formulates the proposed stochastic MILP model for the system restoration problem considered in this paper. Fig.~\ref{fig:Line_Bus_model} shows a sketch of the proposed strategy along with the line and bus models. Define the sets $\mathit{\Pi}$, $\mathcal{N}$ and $\mathcal{T}$, respectively, corresponding to the damage scenarios, the system's buses, and the time periods for the restoration problem. Let subscripts $\pi$, $t$, and $m$ denote a particular scenario, time, and bus, respectively. Each time period has a duration of $\Delta t$ (e.g., 5 minutes) and we consider a horizon lasting a few hours after the disaster. The load at each bus~$m$ has an assigned weight $\omega_{m}$ indicating its importance. The variable $P_{\pi,t, m}^{LC}$ denotes the curtailment of active power load at bus $m$ and time $t$ under scenario $\pi$. The probability of each scenario is denoted by $pr_{\pi}$. The objective~\eqref{eq:objective} minimizes the total unserved energy during the restoration process, weighted by the load's importance:

\begin{figure}[t]
    \centering
    \includegraphics[width=11cm]{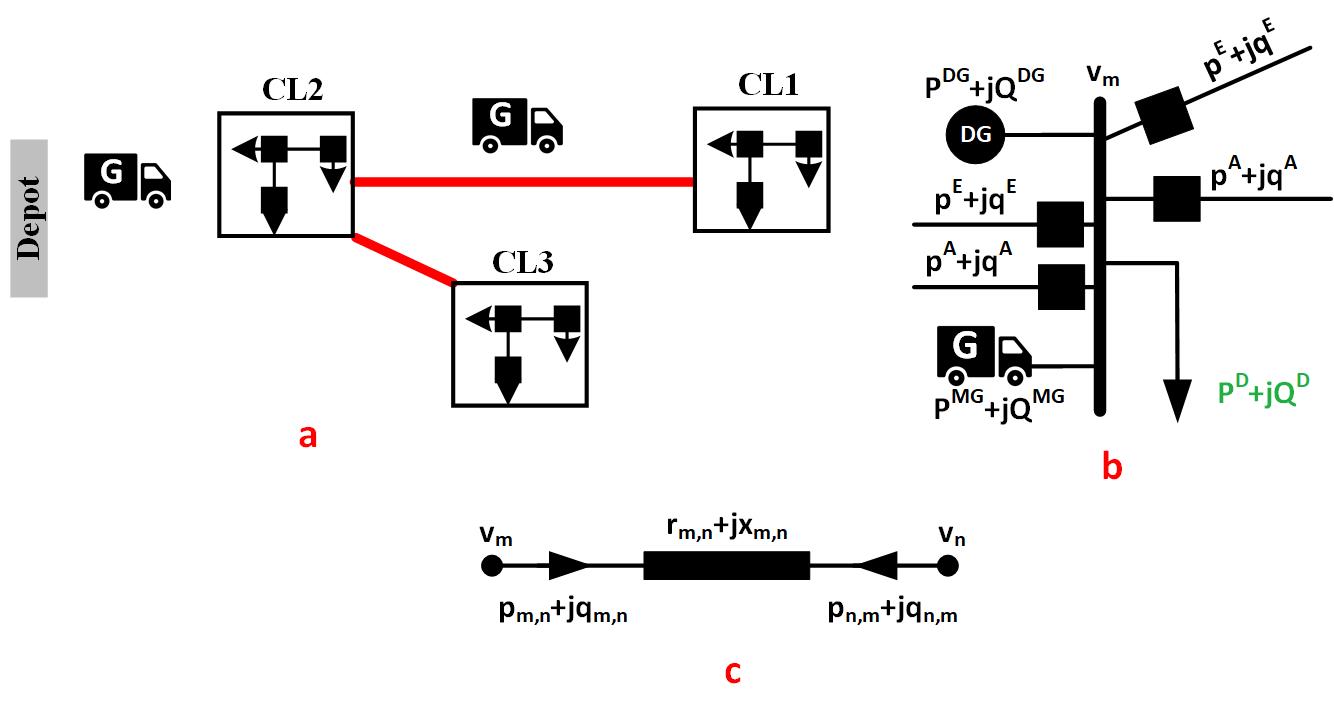}
    \caption{a) Mobile generator and underground line models, b) bus model, and c) line model.}
    \label{fig:Line_Bus_model}
\end{figure}
\begin{equation}
\label{eq:objective}
 \min \sum_{\pi\in \mathit{\Pi}} \sum_{m\in\mathcal{N}} \sum_{t\in\mathcal{T}} \omega_{m} \cdot pr_{\pi} \cdot P_{\pi,m, t}^{L C} \cdot \Delta t
\end{equation}

\subsection{Power Flow}
Let $\mathcal{L}^{E}$, $\mathcal{L}^{A}$, and $\mathcal{L} = \mathcal{L}^{E} \cup \mathcal{L}^{A}$ denote the set of existing lines, the set of added underground lines, and the set of all lines, respectively. 
In this study, we consider a balanced single-phase equivalent model of the distribution systems, but the approach could be generalized to unbalanced three-phase network models as well. Let parameters $P_{t, m}^{D}$ and $Q_{t, m}^{D}$ denote the active and reactive power demand at bus $m$ at time $t$, respectively. We also define the following variables for each time period $t$ and scenario $\pi$:
\begin{itemize}
\item $p_{\pi,t, m,n}^{E}$ and  $q_{\pi,t, m,n}^{E}$: active and reactive  power flows on the existing overhead line connecting buses $m$ and $n$,
\item $p_{\pi,t, m, n}^{A}$ and  $q_{\pi,t, m, n}^{A}$: the active and reactive power flows on the added underground line connecting buses $m$ and~$n$,
\item $P_{\pi,t, m}^{DG}$ and $Q_{\pi,t, m}^{DG}$: active and reactive power of the distributed generator at bus~$m$,
\item $P^{MG}_{\pi,t, m}$ and $Q^{MG}_{\pi,t, m}$: active and reactive power of the mobile generator at bus~$m$.
\end{itemize}
The power balance constraints at each bus $m$ are:
\begin{subequations}
\label{eq:power balance}
\begin{align}
\begin{split} 
  P_{t, m}^{D}+\sum_{\substack{(i,n) \in\mathcal{L}^{E}\\ \textrm{~s.t.~} i=m}} p_{\pi, t, m,n}^{E}+\sum_{n\in\mathcal{N}} p_{\pi, t, m, n}^{A}=P_{\pi, t, m}^{LC}+\\
  P_{\pi, t, m}^{DG} + P^{MG}_{\pi, t, mg m}+\sum_{\substack{(m,i) \in\mathcal{L}^{E}\\ \textrm{~s.t.~} i=n}} p_{\pi, t, n,m}^{E} 
\end{split} \\
\begin{split}
  Q_{t, m}^{D}+\sum_{\substack{(i,n) \in\mathcal{L}^{E}\\ \textrm{~s.t.~} i=m}} q_{\pi, t, m,n}^{E}+\sum_{n\in\mathcal{N}} q_{\pi, t, m, n}^{A}=Q_{\pi, t, m}^{LC}+\\
  Q_{\pi, t, m}^{D G}+ Q^{MG}_{\pi, t, m}+\sum_{\substack{(m,i) \in\mathcal{L}^{E}\\ \textrm{~s.t.~} i=n}} q_{\pi, t, n,m}^{E}
  \end{split} 
\end{align}
\end{subequations}

Equation (\ref{eq:curtailment}) implies that load curtailment at each bus cannot exceed the total demand of the bus. We also consider a constant power factor load curtailment model as indicated by (\ref{eq:power factor}).
\begin{subequations}
\begin{align}
\label{eq:curtailment}
& 0 \leq P_{\pi, t, m}^{LC} \leq P_{t, m}^{D}  \\
\label{eq:power factor}
& Q_{\pi, t, m}^{L C} \cdot {P_{t, m}^{D}}=P_{\pi, t, m}^{L C}  \cdot {Q_{t, m}^{D}}
\end{align}
\end{subequations}

Let $v_{\pi, t, n}$ denote the voltage magnitude at bus $n$ during time~$t$ in scenario~$\pi$. 
$M$ is a big-M constant. Let $r_{m,n}, x_{m,n}$ denote the  resistance and reactance of line $(m,n)$. $\varsigma_{\pi, t, m,n}$ is a binary variable indicating whether the switch on the line $(m,n)$ at time $t$ in scenario $\pi$ is closed. 
$\underline{v}_{m}$ and $\overline{v}_{m}$ are the upper and lower bounds on the voltage magnitude at bus $m$. To obtain a MILP formulation, we use the linearized DistFlow approximation of the power flow equations~\cite{baran1989}, as in other related work, e.g.,~\cite{Taheri2019,Amirioun2018Resilience}. Thus, the voltages of the buses are related as in \eqref{eq:Distflow}, where the big-M method is used to decouple the voltages of two disconnected buses to account for the behavior of switches. 
Furthermore, the voltages of the buses should be within the allowable range as dictated in (\ref{eq:voltage limit}).
\begin{subequations}

\begin{dmath}
\label{eq:Distflow}
  (1-\varsigma_{\pi, t, m,n}) M-\frac{r_{m,n} \cdot p_{\pi, t, m,n}^{E}+x_{m,n} \cdot q_{\pi, t, m,n}^{E}}{v_{1}} 
  \leq v_{\pi, t, n}-v_{\pi, t, m} \leq (\varsigma_{\pi, t, m,n}-1) M-\frac{r_{m,n} \cdot p_{\pi, t, m,n}^{E}+x_{m,n} \cdot q_{\pi, t, m,n}^{E}}{v_{t}} 
\end{dmath}
\begin{equation}
\label{eq:voltage limit}
\underline{v}_{m} \leq v_{\pi, t, m} \leq \overline{v}_{m}    
\end{equation}
\end{subequations}

We linearize the line flow limits specified in terms of apparent power as shown in (\ref{eq:Thermal limits}), where $\bar{S}_{m,n}$ is the capacity of line~$(m,n)$.
\begin{subequations}
\label{eq:Thermal limits}
\begin{align}
 & -\varsigma_{\pi, t, m,n} \cdot \bar{S}_{m,n} \leq p_{\pi, t, m,n}^{E} \leq \varsigma_{\pi, t, m,n} \cdot \bar{S}_{m,n}  \\
 & -\varsigma_{\pi, t, m,n} \cdot \bar{S}_{m,n} \leq q_{\pi, t, m,n}^{E} \leq \varsigma_{\pi, t, m,n} \cdot \bar{S}_{m,n}   \\
 & -\sqrt{2} \varsigma_{\pi, t, m,n} \cdot \bar{S}_{m,n} \leq p_{\pi, t, m,n}^{E}+q_{\pi, t, m,n}^{E} \leq \sqrt{2} \varsigma_{\pi, t, m,n} \cdot \bar{S}_{m,n}  \\
& -\sqrt{2} \varsigma_{\pi, t, m,n} \cdot \bar{S}_{m,n} \leq p_{\pi, t, m,n}^{E}-q_{\pi, t, m,n}^{E} \leq \sqrt{2} \varsigma_{t m,n} \cdot \bar{S}_{m,n}    
\end{align}
\end{subequations}

Let $\alpha_{\pi, t, m}$ be a binary variable indicating whether the distributed generator at bus $m$ is generating power during time $t$ in scenario $\pi$. $\overline{P}_{m}^{DG}$, $\overline{Q}_{m}^{DG}$, $\underline{P}_{m}^{DG}$, and $\underline{Q}_{m}^{DG}$ are the maximum/minimum active/reactive power outputs of the distributed generator at bus $m$. 
Constraint (\ref{eq:DG}) ensures that the active and reactive power outputs of distributed generators are within their allowed ranges. 
\begin{subequations}
\label{eq:DG}
\begin{align}
\alpha_{\pi, t, m} \cdot \underline{P}_{m}^{DG} \leq P_{\pi, t, m}^{DG} \leq \alpha_{\pi, t, m} \cdot \overline{P}_{m}^{DG} \\
\alpha_{\pi, t, m}  \cdot \underline{Q}_{m}^{DG} \leq Q_{\pi, t, m}^{D G} \leq \alpha_{\pi, t, m}  \cdot \overline{Q}_{m}^{DG}    
\end{align}
\end{subequations}

Finally, (\ref{eq:Radial}) enforces a radial configuration of distribution system by using the spanning tree approach wherein each bus except the root bus has either one or zero parent buses~\cite{jabr2012minimum}. In (\ref{eq:Radial}), $\lambda_{\pi, t, m, n}$ is a binary variable indicating whether bus $n$ is the parent of bus $m$ at time $t$ in scenario $\pi$ and $\Psi_{m}$ denotes the set of buses connected to bus $m$ by a line.
\begin{subequations}
\label{eq:Radial}
\begin{align}
& \lambda_{\pi, t, m, n}+\lambda_{\pi, t, n, m}=\varsigma_{\pi, t, m, n}, & \forall n \in \Psi_{m} \\
& \sum_{m \in \Psi_{m}} \lambda_{\pi, t, m, n} \leq 1, & \forall n \in \Psi_{m} \\
& \lambda_{\pi, t, 1, n}=0, & \forall n \in \Psi_{\text {root }}  
\end{align}
\end{subequations}

\subsection{Mobile Generators}
As shown in Fig.~\ref{fig:Line_Bus_model}, mobile generators can be dispatched from depots to distribution buses to supply loads after HILP events. This motivates careful consideration of travel and setup time requirements. In this regard, let $\chi$ and $\Xi$ denote the sets of depots and mobile generators, respectively. $\delta_{\pi, dp,mg, m}$ is a binary variable indicating whether mobile generator $mg$ moves from depot $dp$ to bus $m$ in scenario $\pi$. $\beta_{m}$ is the maximum number of mobile generators that can be installed at bus $m$. Constraints (\ref{eq:mobile1})--(\ref{eq:mobile9}) model mobile generator scheduling after the occurrence of a major disaster. Constraint \eqref{eq:mobile1} indicates that at most $\beta_{m}$ mobile generators can be dispatched to bus $m$.
\begin{equation}
\label{eq:mobile1}
\sum_{mg\in\Xi} \sum_{d p\in\chi} \delta_{\pi, dp,mg, m} \leq \beta_{m}  
\end{equation}
Constraint (\ref{eq:mobile2}) indicates that each mobile generator cannot be dispatched more than once.
\begin{equation}
\label{eq:mobile2}
\sum_{m\in\mathcal{N}} \sum_{d p\in\chi} \delta_{\pi, dp,mg, m} \leq 1 
\end{equation}
Let $\Gamma_{\pi, dp,mg, m}$ denote the arrival time of mobile generator $mg$ from depot $dp$ at bus $m$ in scenario $\pi$. Let $T_{ dp, m}$ denote the time needed for a mobile generator to connect to bus~$m$ when departing from depot $dp$. $\gamma_{\pi, t, mg, m}$ is a binary variable indicating whether a mobile generator arrives at bus~$m$ at time~$t$ in scenario~$\pi$. Constraint~(\ref{eq:mobile3}) models the total time required to connect mobile generators to the distribution buses. 
Constraints (\ref{eq:mobile4}) and (\ref{eq:mobile5}) then ensure consistency among the decision variables related to the connection times. For example, if the arrival time of mobile generator~1 to bus 14 is 20 minutes (i.e., $\Gamma_{\pi, dp, 1, 14}=20$), the binary variable $\gamma_{\pi, t, 1, 14}$ will be one for $t=20$ and zero for $t \neq 20$.
 \begin{align}
\label{eq:mobile3}
\Gamma_{\pi, dp,mg, m}= T_{ dp, m} \times \delta_{\pi,dp,mg, m}  \\
\label{eq:mobile4}
\sum_{t\in\mathcal{T}} t \times \gamma_{\pi, t,mg, m} \geq \sum_{dp\in\chi} \Gamma_{\pi, dp,mg, m} \\
\label{eq:mobile5}
 \sum_{t\in\mathcal{T}} t \times \gamma_{\pi, t,mg, m} \leq \sum_{dp\in\chi} \Gamma_{\pi, dp,mg, m}+1-\varepsilon   
\end{align}
$\kappa_{\pi, t, mg, m}$ is a binary variable indicating whether mobile generator $mg$ is connected to bus $m$ at time $t$ in scenario~$\pi$. Equation (\ref{eq:mobile6}) couples $\gamma_{\pi, t,mg, m}$ and $\delta_{\pi, dp,mg, m}$.
\begin{equation}
\label{eq:mobile6}
\sum_{t\in\mathcal{T}} \gamma_{\pi, t,mg, m}=\sum_{dp\in\chi} \delta_{\pi, dp,mg, m}
\end{equation}
In our model, we consider a time frame during which mobile generators can move from their depot and connect to one bus but cannot subsequently disconnect and move to another bus. This is enforced by (\ref{eq:mobile7}).
\begin{equation}
\label{eq:mobile7}
 \kappa_{\pi, t,mg, m}=\sum_{t=1}^{t} \gamma_{\pi, t,mg, m}   
\end{equation}
$\overline{P}_{mg}$ and $\overline{Q}_{mg}$ are the maximum active and reactive power outputs of mobile generator $mg$, as modeled by (\ref{eq:mobile8}) and (\ref{eq:mobile9}).

\begin{equation}
\label{eq:mobile8}
 0 \leq P ^{MG}_{\pi, t, m} \leq \sum_{mg\in\Xi} \kappa_{\pi, t,mg, m} \times \overline{P}_{mg}    
\end{equation}
\begin{equation}
\label{eq:mobile9}
 0 \leq Q^{MG} _{\pi, t, m} \leq \sum_{mg\in\Xi} \kappa_{\pi, t,mg, m} \times \overline{Q}_{mg}
\end{equation}

\subsection{Underground Lines}
Construction of underground lines can significantly increase the resilience of the power supplied to critical loads after HILP events, especially when installed while considering the capabilities of with mobile generators. In this regard, let 
$\tau_{m, n}$ be a binary variable indicating whether an underground line is constructed between buses~$m$ and~$n$.
The voltages of the underground lines' terminal buses are related by (\ref{eq:undergroundDistFLow}). Constraint (\ref{eq:underground_Servicable}) states that underground lines are only in a serviceable state if the lines are constructed (i.e., $\tau_{m, n}=1$) and the lines' switches are closed (i.e., $\varsigma_{\pi, t, m, n}=1$). 

\begin{subequations}
\begin{align}
\nonumber
& \left(1-\varsigma_{\pi, t, m, n}\right) M-\frac{r_{m, n} \cdot p_{\pi, t, m, n}^{A}+x_{m, n} \cdot q_{\pi, t, m, n}^{A}}{v_{1}} \\ 
& \nonumber \quad\leq v_{\pi, t, n}-v_{\pi, t, m} \\ 
\label{eq:undergroundDistFLow} & \quad\leq\left(\varsigma_{\pi, t, m, n}-1\right) M-\frac{r_{m, n} \cdot p_{\pi, t, m, n}^{A}+x_{m,n} \cdot q_{\pi, t, m, n}^{A}}{v_{1}}\\
\label{eq:underground_Servicable}
& \tau_{m, n} \geq \varsigma_{\pi, t, m, n}
\end{align}
\end{subequations}
The linearized form of the underground lines' flow limits is modeled in (\ref{eq:underground_Linearized}).
\begin{subequations}
\label{eq:underground_Linearized}
\begin{align}
& -\varsigma_{\pi, t, m, n} \cdot \overline{S}_{m, n} \leq p_{\pi, t, m, n}^{A} \leq \varsigma_{\pi, t, m, n} \cdot \overline{S}_{m, n}  \\
& -\varsigma_{\pi, t, m, n} \cdot \overline{S}_{m, n} \leq q_{\pi, t, m, n}^{A} \leq \varsigma_{\pi, t, m, n} \cdot \overline{S}_{m, n} \\
& -\sqrt{2} \varsigma_{\pi, t, m, n} \cdot \overline{S}_{m, n} \leq p_{\pi, t, m, n}^{A}+q_{\pi, t, m, n}^{A} \leq \sqrt{2} \varsigma_{\pi, t, m, n} \cdot \overline{S}_{m, n} \\
& -\sqrt{2} \varsigma_{\pi, t, m, n} \cdot \overline{S}_{m, n} \leq p_{\pi, t, m, n}^{A}-q_{\pi, t, m, n}^{A} \leq \sqrt{2} \varsigma_{\pi, t, m, n} \cdot \overline{S}_{m, n}    
\end{align}
\end{subequations}
Let 
$\ell_{m n}$ denote the length of the line between buses~$m$ and~$n$.  $C^{A}$, $C^{\text {RCS }}$, and $C^{\text {invest }}$ are the per-length cost of constructing underground lines, the cost of installing remote-controlled switches, and the total budget of the distribution utility, respectively. $\eta^{A}$ is the maximum allowed number of underground lines.
Constraint (\ref{eq:underg_3}) ensures that the total investment cost of underground lines and remote-controlled switches is no greater than the total budget of the distribution utility. Also, (\ref{eq:underg_4}) states that the number of constructed underground lines must be within the specified limit.
\begin{subequations}
\label{eq:underg}
\begin{align}
\label{eq:underg_3}
& \sum_{\substack{(m,n) \in\mathcal{L}^{A}}}\left(\frac{1}{2} \ell_{m n}^{A} \cdot C^{A}+C^{\text {RCS }}\right) \cdot \tau_{m, n} \leq C^{\text {invest }} \\
\label{eq:underg_4}
& \sum_{\substack{(m,n) \in\mathcal{L}^{A}}} \tau_{m, n} / 2 \leq \eta^{A}    
\end{align}
\end{subequations}

\subsection{Final Model}

The final model is summarized as follows:
\begin{equation}
\label{eq:final model}
 \min ~ (\ref{eq:objective})
  \textrm{~s.t.~} (\ref{eq:power balance})-(\ref{eq:underg})   
\end{equation}


\begin{table}[!t]

\caption{Computational Scaling of the Stochastic MILP Model}
\label{Computational Scaling}
\centering

\begin{tabular}{|c|c|c|}
\hline
\textbf{Num. of} & \textbf{Computational scaling} & \textbf{IEEE 123-Bus}\\
\hline

\thead{Binary\\ variables}& $\thead{|\Pi||T|( 2|L|+3|N|)\\ +|\Pi||\Xi||N|+|N|^{2}+|\eta|}$ & 324,554 \\

\thead{Continuous\\ variables}&$\thead{2|\Pi||T|(|N|+|L|+|\eta|+|DG|)\\ +3|\Pi||N||\Xi|}$& 279,780 \\

Constraints& $\thead{|T||\Pi|(7|N|+5|L|+3|\Xi||N|+7|\eta| \\ +2|DG|+2|L|+1+2|N|) \\ +2|\Pi||\Xi|  +4|\Pi||N||\Xi|+|\eta|+2}$& 1,908,447\\
\hline
\end{tabular}
\end{table}

Table~\ref{Computational Scaling} describes the computational scaling of the proposed model. Observe that the number of variables has a linear relationship with the number of time periods and the number of scenarios. 

\section{Results and Discussion}
\label{sec:Results and Discussion}
In this section, the model from Section~\ref{sec:Mathematical Formulation} is applied to the IEEE~123-bus test system to demonstrate the effectiveness of the proposed approach.
\subsection{IEEE 123-Bus Test System}
\begin{figure}[htp]
    \centering
    \includegraphics[width=11cm]{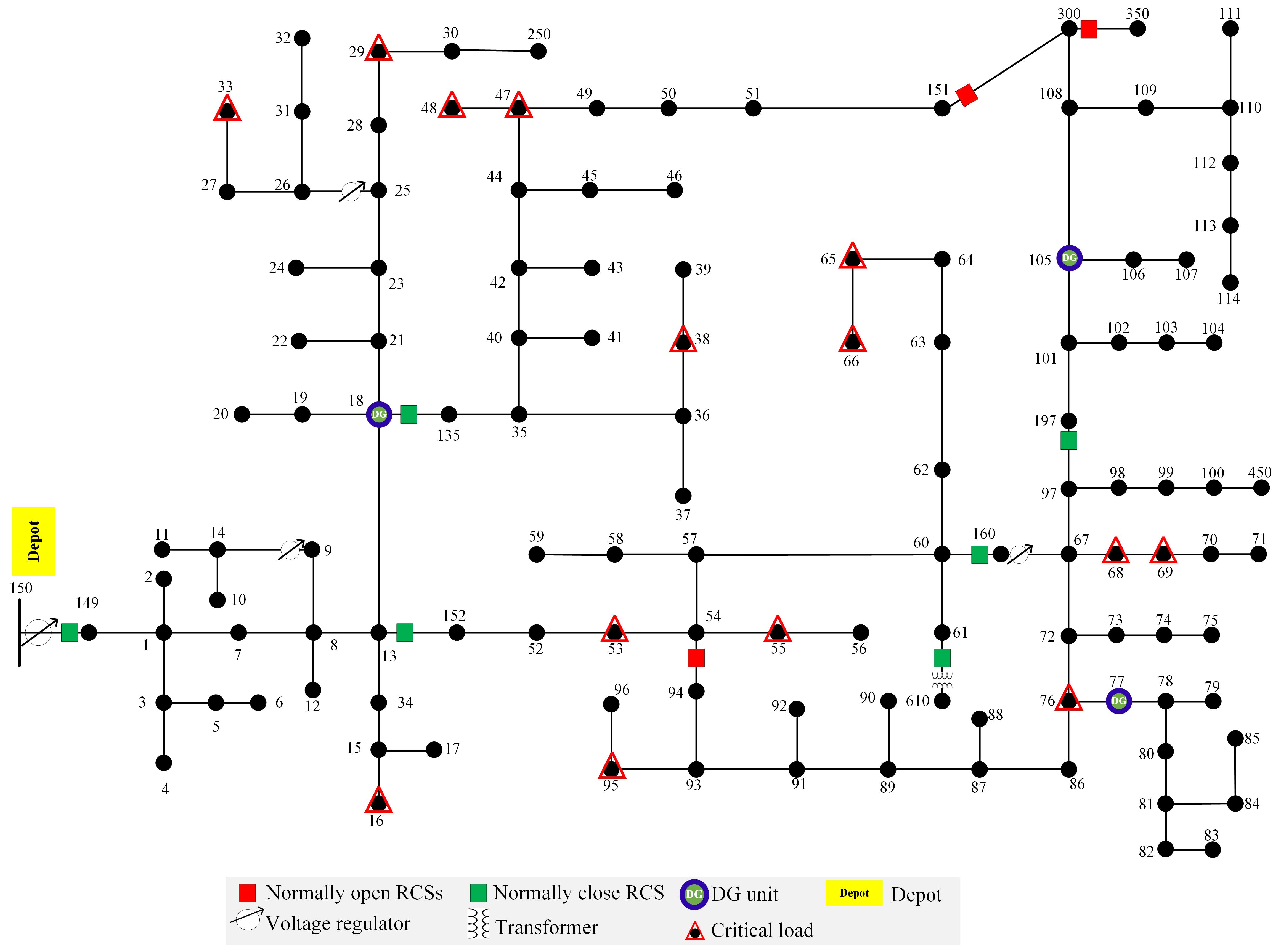}
    \caption{Modified IEEE 123-bus test system.}
    \label{fig:Modified IEEE 123-bus test system}
\end{figure}

This system is operated at a nominal voltage of 4.16 kV with total active and reactive power demand of 3.49~MW and 1.92 MVAr,~respectively. The modified test system is shown in Fig.~\ref{fig:Modified IEEE 123-bus test system}. Three distributed generators are installed in the test system with the parameters given in Table~\ref{Parametersof DGs}. Also, five mobile generators are available at the depot location with the parameters shown in Table~\ref{Parametersof mobile generators}. Here, 14 of the loads are classified as ``critical'' as shown by the triangles in Fig.~\ref{fig:Modified IEEE 123-bus test system}. The underground distribution lines are equipped with remote-controlled switches. The costs of
constructing underground distribution lines and installing remote-controlled switches are given in Table~\ref{The Investment Cost}. 

We solved the stochastic MILP problem described in Section~\ref{sec:Mathematical Formulation} to a 1\% optimality gap using the Gurobi~9.5.0 solver on a computing node of the Partnership for an Advanced Computing Environment (PACE) cluster at the Georgia Institute of Technology~\cite{PACE}. This computing node has a 24-core CPU and 32~GB of RAM. We implement the model in GAMS~38.1~\cite{mccarl2014mccarl}.

\begin{table}[!t]

\caption{Parameters of Installed Distributed Generators (DG) in the IEEE 123-Bus Test System}
\label{Parametersof DGs}
\centering
\begin{tabular}{|c|c|c|c|c|c|c}
\hline
\textbf{DG} & \textbf{Bus} & $\mathbf{\overline{P}^{DG}_{m}}$ & $\mathbf{\underline{P}^{DG}_{m}}$ & $\mathbf{\overline{Q}^{DG}_{m}}$ & $\mathbf{\underline{Q}^{DG}_{m}}$\\
\hline

DG1& 18& 400~kW & 0~kW &300~kVAr& -300~kVAr\\

DG2 &77& 700~kW& 0~kW& 600~kVAr& -600~kVAr\\

DG3& 105& 700~kW& 0~kW& 600~kVAr& -600~kVAr\\
\hline
\end{tabular}
\end{table}

\begin{table}[!t]

\caption{Parameters of Available Mobile Generators}
\label{Parametersof mobile generators}
\centering
\begin{tabular}{|c|c|c|}
\hline
\textbf{Mobile Generators} & $\mathbf{\overline{P}_{MG}}$ & $\mathbf{\overline{Q}_{MG}}$ \\
\hline
MG1& 200~kW & 150~kVAr\\
MG2 &300~kW& 200~kVAr\\
MG3 &500~kW& 400~kVAr\\
MG4 &650~kW& 550~kVAr\\
MG5 &700~kW& 600~kVAr\\
\hline
\end{tabular}
\end{table}

\begin{table}[!t]

\caption {Investment Costs}
\label{The Investment Cost}
\centering

\begin{tabular}{|c|c|}
\hline
\textbf{Investment} & \textbf{Cost}\\
\hline
Underground line construction & \$1M/mile~\cite{wang2015networked}\\
Remote-controlled switches & \$15k~\cite{taheri2019hardening}\\
\hline
\end{tabular}
\end{table}

\subsection{Simulation Results}
We next describe the results obtained from solving the MILP model (\ref{eq:final model}) to determine locations for adding new underground lines while jointly modeling many disaster scenarios. For each scenario, this model considers the optimal dispatch of the mobile generators when determining where to add new underground lines. The results show that this holistic modeling approach in the planning phase improves the system's restoration performance during the operational phase.
\subsubsection{Planning}
To find the optimal locations for constructing
new underground distribution lines, we solve the proposed MILP model (\ref{eq:final model}) with 20 scenarios. To capture the uncertainties associated with damage to distribution lines, we first generated 1000 damage scenarios by a Monte Carlo sampling technique along with fragility curves~\cite{Taheri2019} and then reduced these to 20 scenarios using GAMS' \texttt{scenred} tool. 

The optimal locations for varying numbers of underground lines, their associated costs, and computational times are given in Table~\ref{Optimal Distribution Line Undergrounding}. As an example, with a \$1 million budget, the solution constructs five underground lines with a total length of 4100~feet (1250~meters) to link certain critical loads such that the operator can simultaneously re-energize them after a disaster. 
Consider, for instance, a scenario where buses 29 and 47 are isolated due to a natural disaster. In this situation, the system operator can restore both loads simultaneously by dispatching one mobile generator to the location of the nearest bus instead of dispatching two mobile generators separately to the locations of buses 29 and 47.

\subsubsection{Operation}
\begin{figure}[t]
    \centering
    \includegraphics[width=11cm]{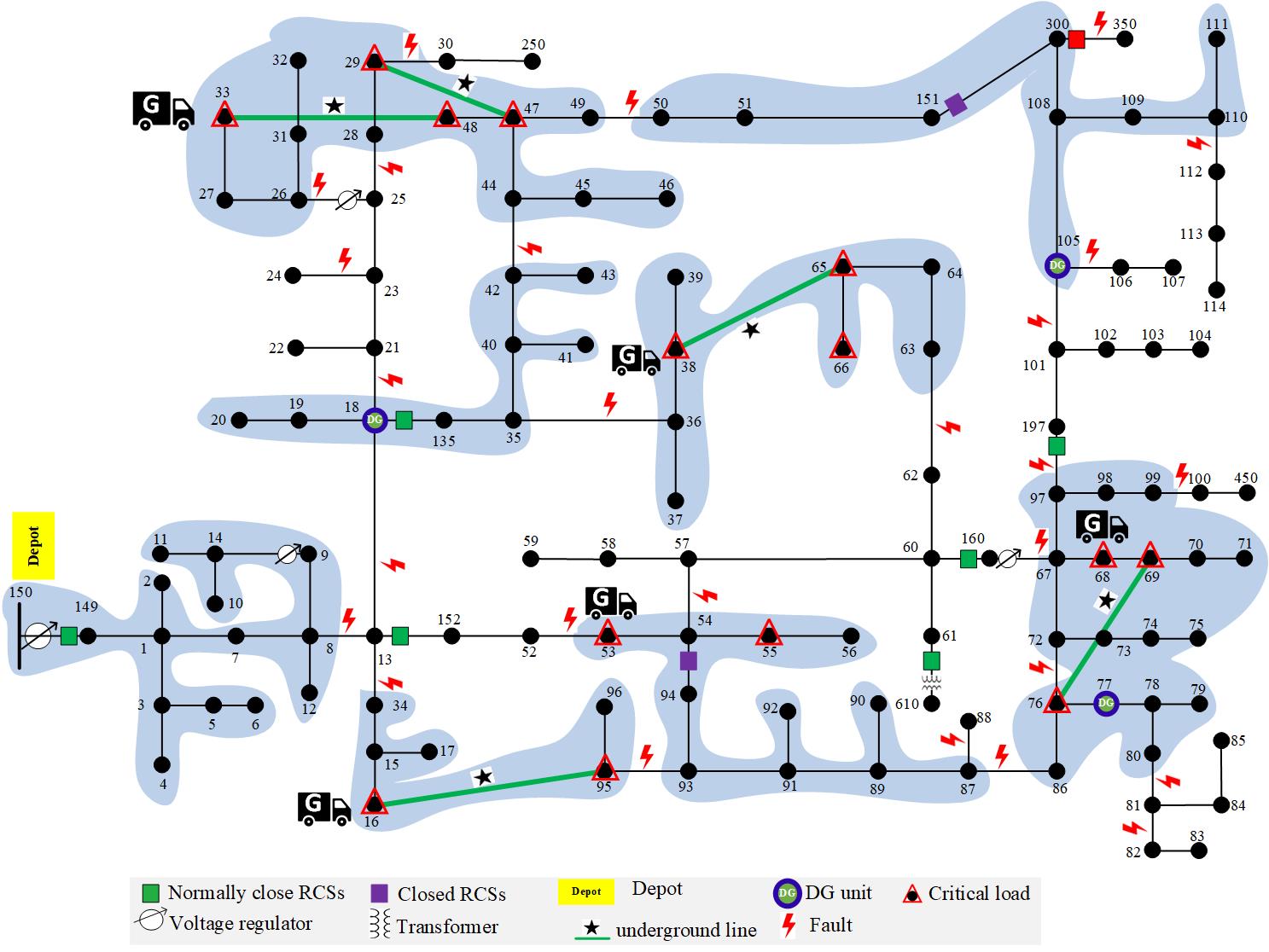}
    \caption{Optimal dispatch of mobile generators in the IEEE 123-bus test system.}
    \label{fig:Optimal dispatch of mobile generators in IEEE 123-bus test system}
\end{figure}

As a representative example, we next focus on the case in Table~\ref{Optimal Distribution Line Undergrounding} where five underground distribution lines are constructed; see the green lines marked with a star in Fig.~\ref{fig:Optimal dispatch of mobile generators in IEEE 123-bus test system} for their locations. Note that the distribution system is operated radially by appropriately configuring the switches on the newly added underground lines. For a representative scenario where 27 of the lines have been damaged due to a natural disaster (marked with red lightening bolts in Fig.~\ref{fig:Optimal dispatch of mobile generators in IEEE 123-bus test system}), we use the proposed model (\ref{eq:final model} with fixed locations for underground lines and this particular damage scenario to optimize the dispatch of mobile generators and the system configuration via the switch statuses in order to restore service to the maximum extent possible. 
Fig.~\ref{fig:Optimal dispatch of mobile generators in IEEE 123-bus test system} visualizes the solution, including the dispatch of available mobile generators, and Table~\ref{Optimal Dispatch of mobile generators} provides the optimal dispatch of mobile generators and their associated arrival times. We observe that the simultaneous utilization of mobile generators and reliable links between critical loads leads to wide-area energization of the distribution system after the disaster. The mobile generators are directly connected to critical loads and also supply adjacent high-priority customers either through available overhead lines or the newly constructed underground lines. Therefore, the model dispatches the mobile generators to restore the critical loads as fast as possible. For instance, mobile generator~1 is dispatched to both energize the critical load at bus~16 and simultaneously serve the critical load at bus~95. The other mobile generators similarly exploit the previously existing and newly built lines to energize other critical loads. 


\begin{table}[!t]

\caption{Optimal Distribution Line Undergrounding
}
\label{Optimal Distribution Line Undergrounding}
\centering

\begin{tabular}{|c|p{2.7cm}|c|c|c|c}
\hline
\textbf{Num.} & \textbf{Underground Lines} & \textbf{Length (ft)} & \textbf{Cost} & \textbf{Time (sec)} \\ 
\hline

6 &16-55, 53-95, 29-47, 33-48, 38-65, 69-76 & 4900& \$1,108k& 1799\\ 

\textbf{5}& \textbf{16-95, 29-47, 33-48, 38-65, 69-76}& \textbf{4100}& \textbf{\$926.5k}& \textbf{1879}\\ 
4& 29-47, 33-48, 38-65, 69-76& 3100& \$707.1k&2233\\ 
3& 29-47, 33-48, 69-76& 2200& \$506.7k&2492\\ 
2& 29-47, 38-65& 1500& \$344.1k&1062\\ 
1& 29-47& 700& \$162.6k& 334\\ 
\hline
\end{tabular}
\end{table}

\begin{figure}[t]
    \centering
    \includegraphics[width=11cm]{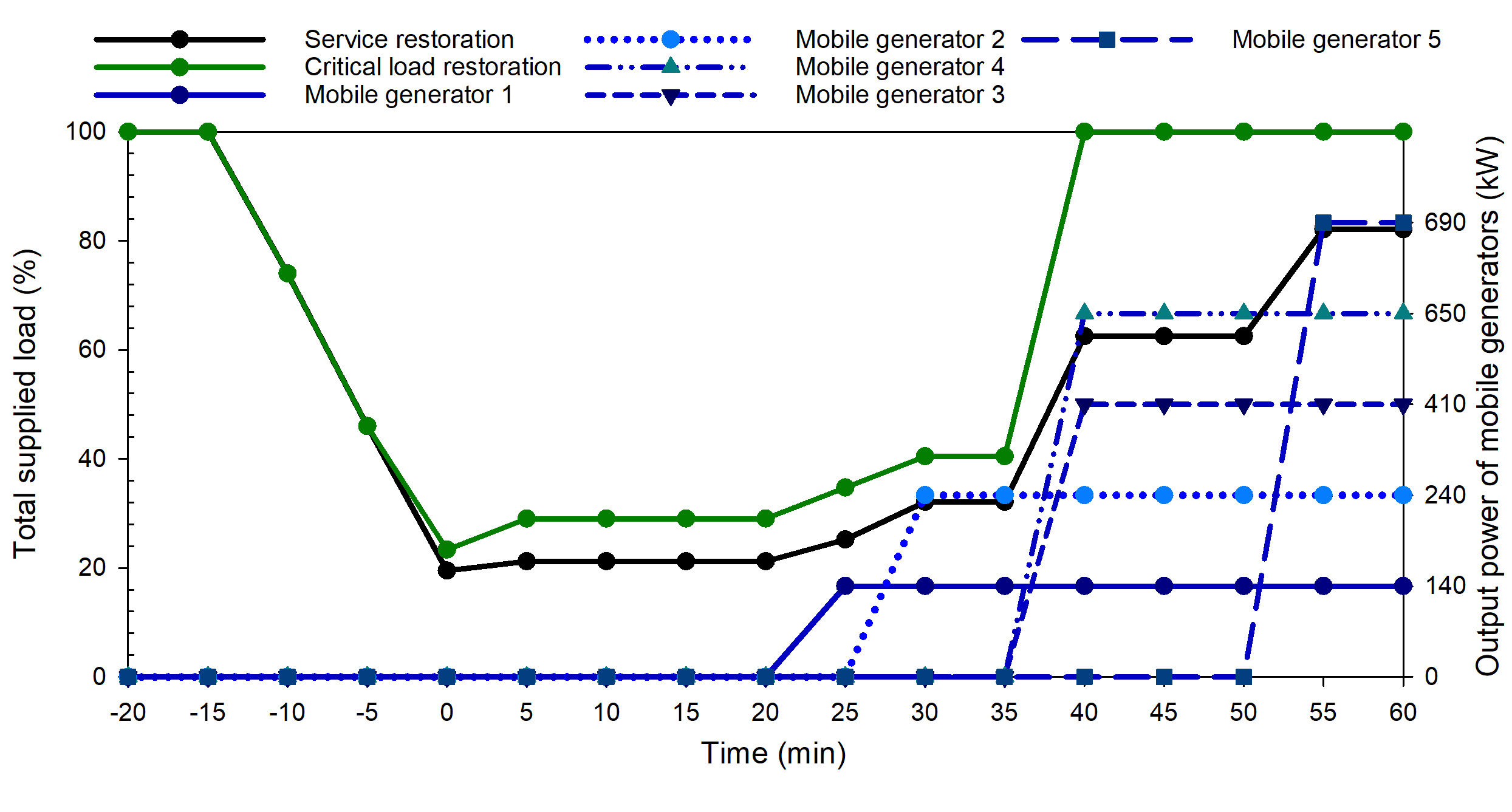}
    \caption{Optimal service restoration in the IEEE 123-bus test system.}
    \label{fig:Optimal service restoration in IEEE 123-bus test system}
\end{figure}

Fig.~\ref{fig:Optimal service restoration in IEEE 123-bus test system} shows a temporal representation of the service restoration process, including the total supplied load, the supplied critical load, and the power outputs of the mobile generators. As more mobile generators reach their destinations and connect to the system, the total supplied load increases from 19.5\% to 82.1\% at the end of the restoration process. Moreover, as shown in by the utilization rates (ratio of power output to generation capacity) in Table~\ref{Utilization Rate of mobile generators}, the solution effectively uses the capabilities of the mobile generators. The solution energizes all critical loads within the first 40 minutes of the restoration process.

\begin{figure}[t]
    \centering
    \includegraphics[width=11cm]{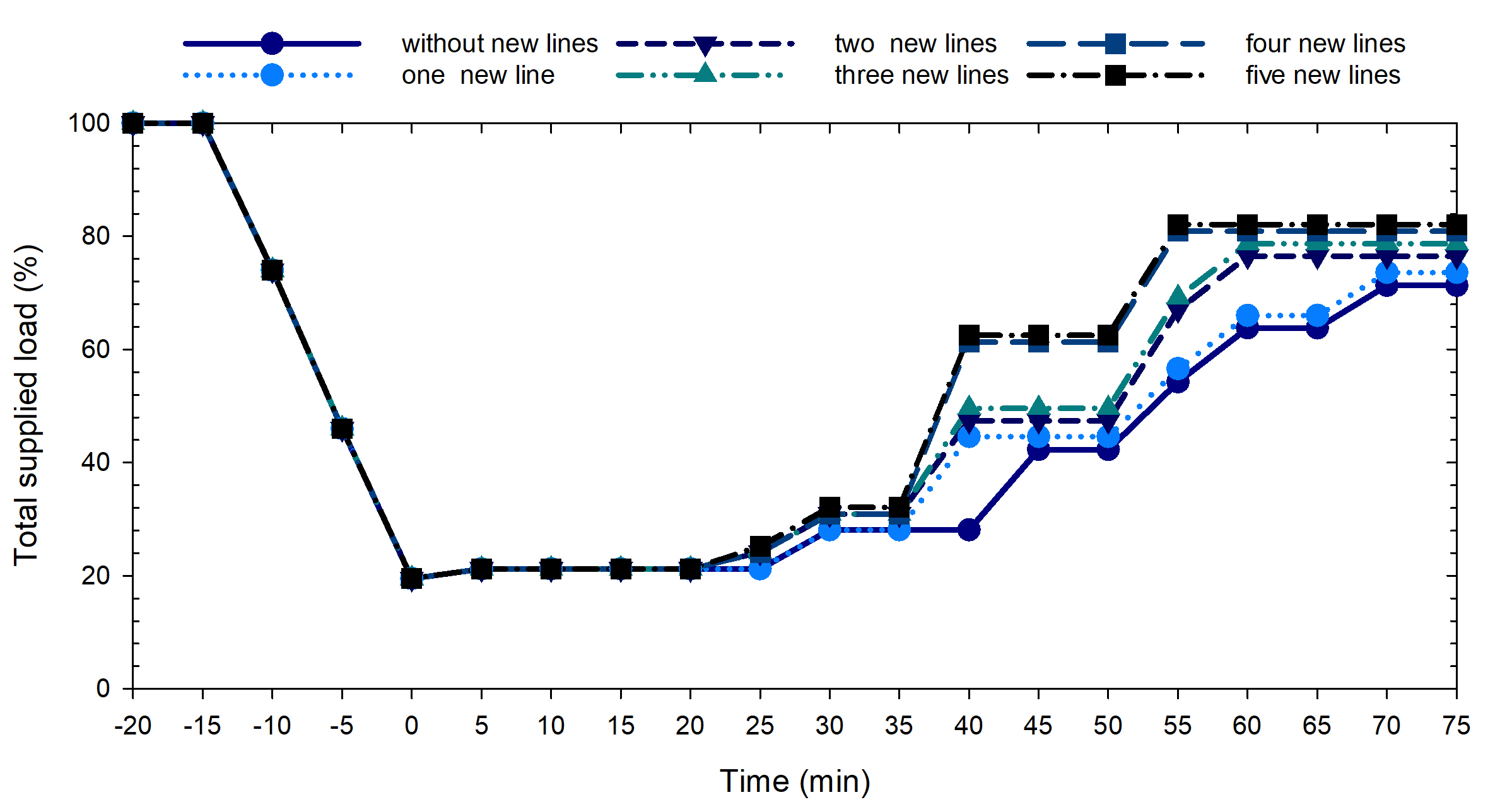}
    \caption{Load supplied with and
without underground lines.}
    \label{fig:Comparison of the total supplied load level}
\end{figure}

Fig.~\ref{fig:Comparison of the total supplied load level} plots the load supplied during the restoration processes corresponding to investment plans with differing numbers of underground lines. When five new lines are constructed, the total energy supplied within the first two hours increases by 3.985~MWh (57.1\%) compared to the case without any new lines. The results indicate that the presence of underground distribution lines increases the utilization rate of mobile generators by transmitting some of their output power to the neighboring critical loads.

\begin{figure}[t]
    \centering
    \includegraphics[width=11cm]{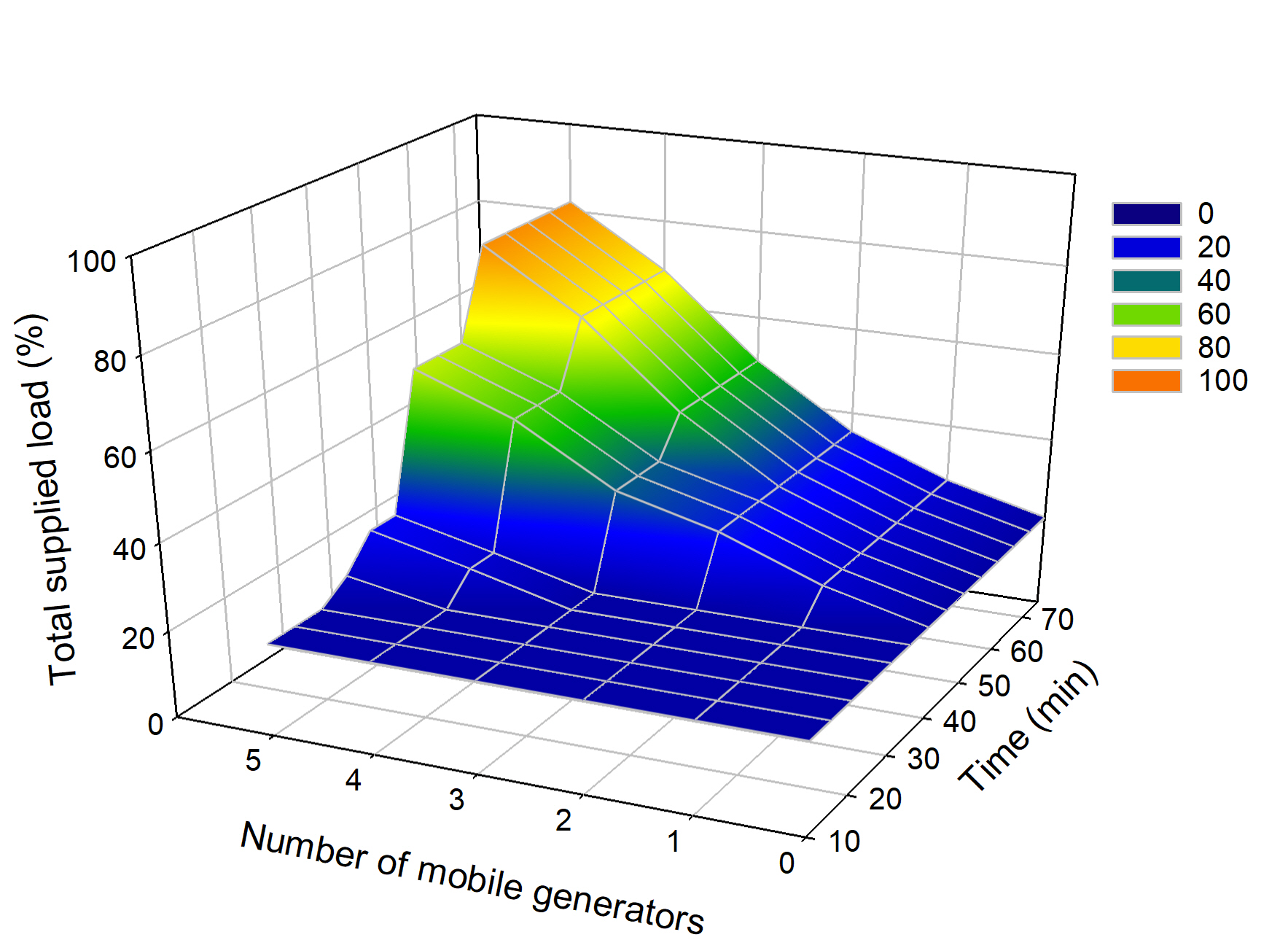}
    \caption{Load supplied with various numbers of mobile generators.}
    \label{fig:The total supplied load level of the system with various number of mobile generators}
\end{figure}

 To assess the value of additional mobile generators, Fig.~\ref{fig:The total supplied load level of the system with various number of mobile generators} visualizes the load served during restoration processes that use various numbers of mobile generators. For varying numbers of available mobile generators in the system, the total load supplied increases by 1.7\%, 7.4\%, 16\%, 30.1\%, 49\%, and 62.6\% for zero, one, two, three, four, and five available mobile generators, respectively.

\begin{figure}[t]
    \centering
    \includegraphics[width=11cm]{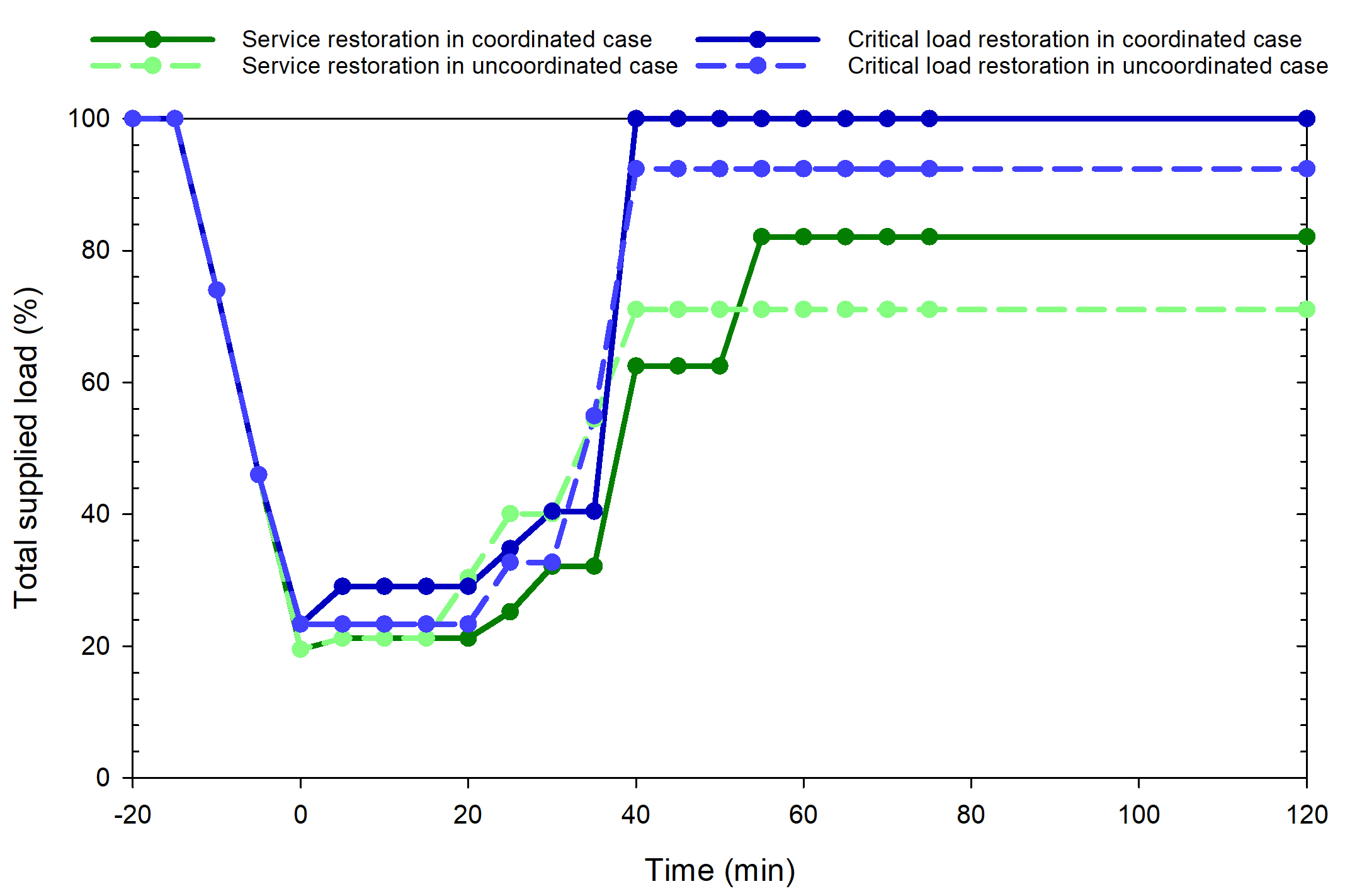}
    \caption{Load supplied when undergrounding decisions are made while considering mobile generators (dark solid lines) and not considering mobile generators (light dashed lines).}
    \label{fig:Zero-MG}
\end{figure}

We finally show that long-term line undergrounding decisions which are cognizant of short-term mobile generator deployments yield superior results relative to undergrounding decisions made without considering mobile generators. To do so, we consider an alternative ``uncoordinated'' approach which sets the number of mobile generators to zero in the stochastic MILP model and computes the best locations for five new underground lines (within the budget constraints) for the specified disaster scenarios. Then, by fixing the locations of these underground lines, the model is executed with five mobile generators for the same disaster scenario as analyzed previously. The restoration process for the total load and the critical load for this case are depicted in Fig.~\ref{fig:Zero-MG}. Observe that the total energy supplied to critical loads in this uncoordinated case decreases by 12.53\% compared to the coordinated case in Fig.~\ref{fig:Optimal service restoration in IEEE 123-bus test system}. Also, the final amounts of total load supplied (71.06\%) and critical load supplied (92.42\%) in the uncoordinated case are less than the corresponding values (82.1\% and 100\%, respectively) for the coordinated case in Fig.~\ref{fig:Optimal service restoration in IEEE 123-bus test system}. Moreover, as shown in Table \ref{Utilization Rate of mobile generators}, the mobile generators in the coordinated case have a higher overall utilization rate compared to the uncoordinated case (90.67\% versus 76.8\%). Thus, considering line undergrounding and mobile generators simultaneously in the proposed model results in better service restoration performance in the aftermath of a natural disaster.


\begin{table}[!t]

\caption {Optimal Dispatch of Mobile Generators}
\label{Optimal Dispatch of mobile generators}
\centering

\begin{tabular}{|c|c|c|}
\hline
\textbf{Mobile Generator} & \textbf{Bus} & \textbf{Arrival Time (min)} \\
\hline
MG1& 16& 25\\
MG2& 53& 30\\
MG3& 38& 40\\
MG4& 33& 40\\
MG5& 68& 55\\
\hline
\end{tabular}
\end{table}




\begin{table}[!t]
\caption {Utilization Rate of Mobile Generators}
\label{Utilization Rate of mobile generators}
\centering
\begin{tabular}{|c|c|c|}
\hline
\multicolumn{3}{|c|}{\textbf{Mobile Generators Utilization Rate}}\\
\hline
\textbf{Mobile Generator} & \textbf{Coordinated case} & \textbf{Uncoordinated case}\\
\hline
MG1 &  70\%  & 40\%\\
MG2 &  80\% & 100\%\\
MG3 & 82\% &  68\%\\
MG4 &  100\%& 78.5\%\\
MG5 &  98.6\%& 82.1\%\\
\hline
Total &  90.64\%& 76.8\%\\
\hline
\end{tabular}
\end{table}

\section{Conclusion}
\label{sec:Conclusion}
To provide critical loads with a highly reliable power supply, this paper proposes a MILP model that considers both the deployment of mobile generators and the optimal locations to underground lines in order to provide utilities with 1) flexible and reliable power sources and 2) resilient connections between locations with critical loads. 
The proposed model has been evaluated using a standard test system and the simulation results confirm the effectiveness of these measures in coping with extreme events.

In future work, we plan to use a more realistic transportation system model for mobile generators that includes, for instance, the potential for traffic interruptions. Finally, another direction for extending the work is to consider a nonlinear representation of the power flow equations, ideally a three-phase unbalanced power flow model, to more accurately represent the behavior of heavily stressed systems.





\bibliography{mybibfile}

\end{document}